\documentclass{article}

\usepackage{microtype}
\usepackage{graphicx}
\usepackage{subfigure}
\usepackage{booktabs} % for professional tables
\usepackage{latexsym}
\usepackage{bbm}
\usepackage{caption}
\usepackage{etoolbox} % for patching caption command
\usepackage{booktabs}
\usepackage{adjustbox}
\usepackage{float}

\makeatletter
\patchcmd{\@makecaption}
  {\centering}
  {\raggedright}
  {}{}
\makeatother

\usepackage{hyperref}

\usepackage[accepted]{icml2025}

\usepackage{amsmath}
\usepackage{amssymb}
\usepackage{mathtools}
\usepackage{amsthm}

\usepackage[capitalize,noabbrev]{cleveref}

\theoremstyle{plain}

\theoremstyle{definition}

\theoremstyle{remark}

\usepackage[textsize=tiny]{todonotes}

\icmltitlerunning{Pi-SAGE: Permutation-invariant surface-aware graph encoder for binding affinity prediction}

\begin{document}

\twocolumn[
\icmltitle{Pi-SAGE: Permutation-invariant surface-aware graph encoder for binding affinity prediction}




\begin{icmlauthorlist}
\icmlauthor{Sharmi Banerjee}{amz}
\icmlauthor{Mostafa Karimi}{amz}
\icmlauthor{Melih Yilmaz}{amz}
\icmlauthor{Tommi Jaakkola}{mit}
\icmlauthor{Bella Dubrov}{amz}
\icmlauthor{Shang Shang}{amz}
\icmlauthor{Ron Benson}{amz}
%\icmlauthor{Firstname8 Lastname8}{sch}
%\icmlauthor{Firstname8 Lastname8}{yyy,comp}
\end{icmlauthorlist}

\icmlaffiliation{amz}{Amazon, Seattle, WA, USA}
\icmlaffiliation{mit}{Massachusetts Institute of Technology, Cambridge, MA,
USA}
%\icmlaffiliation{sch}{School of ZZZ, Institute of WWW, Location, Country}

\icmlcorrespondingauthor{Sharmi Banerjee}{sharmiba@amazon.com}
%\icmlcorrespondingauthor{Firstname2 Lastname2}{first2.last2@www.uk}

\icmlkeywords{Machine Learning, ICML}

\vskip 0.3in
]



\printAffiliationsAndNotice{}  % leave blank if no need to mention equal contribution

\begin{abstract}
Protein surface fingerprint encodes chemical and geometric features that govern protein–protein interactions and can be used to predict changes in binding affinity between two protein complexes. Current state-of-the-art models for predicting binding affinity change, such as GearBind, are all-atom based geometric models derived from protein structures. Although surface properties can be implicitly learned from the protein structure, we hypothesize that explicit knowledge of protein surfaces can improve a structure based model's ability to predict changes in binding affinity. To this end, we introduce Pi-SAGE, a novel Permutation-Invariant Surface-Aware Graph Encoder. We first train Pi-SAGE to create a protein surface codebook directly from the structure and assign a token for each surface exposed residue. Next, we augmented the node features of the GearBind model with surface features from domain adapted Pi-SAGE to predict binding affinity change on the SKEMPI dataset. We show that explicitly incorporating local, context-aware chemical properties of residues enhances the predictive power of all-atom graph neural networks in modeling binding affinity changes between wild-type and mutant proteins.
\end{abstract}

\section{Introduction}
A protein's surface encodes critical chemical and geometric fingerprints such as charge, shape, and hydrogen bond interactions that enables tasks like identifying active binding sites, designing proteins with specific properties, predicting ligand–protein binding affinity \citet{gainza2020deciphering,song2024surfpro, lee2023shapeprot, somnath2021multi} and so on. The seminal MaSIF paper \citet{gainza2020deciphering} demonstrated that these fingerprints can be extracted from protein structures and used efficiently in downstream tasks such as binding site prediction, protein–ligand interaction modeling, and binding site search in protein complexes. Subsequently studies have reinforced the importance of explicitly modeling surface-level chemical and geometric features to perform tasks such as protein function prediction \citet{somnath2021multi} and surface property guided protein design \citet{lee2023shapeprot, song2024surfpro}.

In parallel, recent studies have proposed structure-aware protein language models (PLMs) by incorporating 3D structural information and showed that explicit structure information combined with sequence information improves performance across various predictive tasks. \citet{su2023saprot,li2024prosst}. %In ProstT5 \citet{heinzinger2024bilingual} the authors showed that by learning protein structures from FoldSeek tokens the model performed better than the SOTA sequence-only models on predicting binding and conserved residues. SaProt \citet{su2023saprot} created a structure-aware vocabulary combining residue tokens with structure tokens (derived from Foldseek) and then trained a general-purpose PLM on 40 million protein sequences and structures. 
Most notably, GearBind \citet{cai2024pretrainable}, an all-atom geometric neural network model outperformed SOTA models on the binding affinity prediction problem. 

Motivated by these advances, we hypothesize that while protein surface information may be implicitly learned from structural models, explicitly modeling the surface provides additional inductive bias—especially for tasks such as predicting binding affinity changes at the protein–protein interface. %explicitly modeling the surface will boost a model's ability to predict binding affinity changes where the mutations to a protein complex happens at the surface interface of the binding partners. 
To test our hypothesis, we propose Pi-SAGE: a Permutation-invariant Surface-Aware Graph Encoder that explicitly creates a vocabulary of protein surface from local geometric and chemical features of residues. Pi-SAGE is pre-trained in two stages: first on the ~200k RCSB PDB database \citet{burley2023rcsb} to capture general surface representations, and then on the SKEMPI dataset \citet{jankauskaite2019skempi}, which contains 6k binding affinity data for wild-type and mutated protein complexes. The model learns from residue graphs, where each node encodes a local surface ``snippet" constructed from neighboring atoms’ geometric and chemical properties. We train GearBind \citet{cai2024pretrainable}, all-atom model by augmenting the one-hot residue features with these surface-aware features and demonstrate improved accuracy in predicting $\Delta \Delta G$ (binding affinity changes). Pi-SAGE outperforms both large-scale sequence-based models that attempt to learn structure implicitly as well as existing structure-aware models. %We pre-trained Pi-SAGE in two stages, first on the RSCB database \citet{burley2023rcsb} and then on the SKEMPI database \citet{jankauskaite2019skempi} to create a protein surface codebook by learning from residue graphs where nodes represent surface snippets containing local chemical and geometric features form neighboring atoms. Next, we trained the GearBind \citet{cai2024pretrainable} all-atom model and showed that augmenting the protein structure graph with surface-aware features improves the model's ability to predict $\Delta \Delta G$. We also showed that Pi-SAGE outperforms both very large sequence-only models that learn about structure implicitly as well as structure-aware models. 
In summary, our main contributions are as follows.
\begin{itemize}
  \item We introduce a novel surface-aware vocabulary that builds a protein surface codebook from local geometric and chemical residue-level features.
  \item We pre-train Pi-SAGE in two stages—on the RCSB PDB and SKEMPI databases—to capture both general and task-specific surface information. %Trained the Pi-SAGE model in two stages, first on the RSCB database, and then on SKEMPI database containing wild type and mutated protein complexes.
  \item We empirically demonstrate that augmenting the GearBind model with explicit surface features improves its ability to predict binding affinity changes ($\Delta \Delta G$), outperforming both sequence-only and existing structure-based approaches. %Trained the SOTA GearBind model with augmented surface features and showed that explicit protein surface information improves the already powerful all-atom model's ability to predict $\Delta \Delta G$.
\end{itemize}

\begin{figure*}
    \centering
    \includegraphics[width=0.5\linewidth]{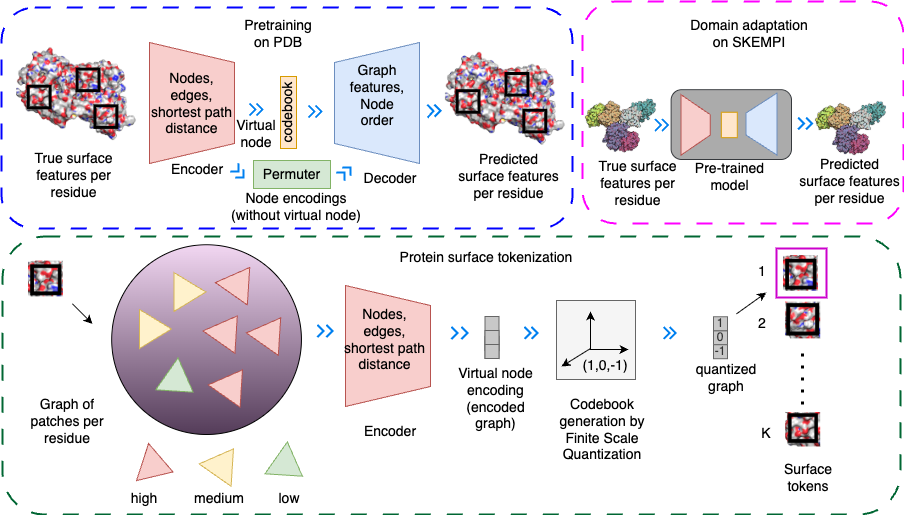}
    \caption{Pi-SAGE is trained in two stages, 1) on the RSCB database with ~200K protein structures (top left), 2) on SKEMPI database with 6k wild type and mutated complexes from 340 unique protein complexes (top right). Each residue is represented by a graph of high, medium and low patches that are first fed to the encoder, followed by the quantizer that creates a codebook for the residue (bottom).}
    \label{fig:flowchart}
\end{figure*}

\section{Related Work}

In recent years, general protein models trained on millions of sequences have rapidly advanced, with most adopting the transformer architecture \citep{vaswani2017attention, rao2020transformer, elnaggar2021prottrans, madani2020progen}. These models, trained on masked language modeling (MLM), have shown that protein structure can be implicitly learned and applied to downstream tasks like contact map and secondary structure prediction, as well as solubility and cellular localization \citep{rao2020transformer, elnaggar2021prottrans}. Model capacity and pretraining strategies have expanded, including span masking in ProtT5 (3B/11B) \citep{elnaggar2021prottrans}, blank-filling in xTrimoPGLM (100B) \citep{chen2024xtrimopglm}, and multi-task learning in AminoBERT \citep{bouatta2022single}.
 
The release of 200M predicted structures by AlphaFold DB \citep{varadi2022alphafold} spurred development of structure-aware models such as ProtT5-XL-UniRef50-Structure \citep{heinzinger2023bilingual} and SaProt \citep{su2023saprot}, which combine sequence and structure inputs and outperform sequence-only models on tasks like contact prediction, thermostability, and protein–protein interaction (PPI) prediction \citep{meier2021language, xu2022peer}.

In parallel, surface-based representations have gained traction due to their ability to capture functionally relevant chemical and geometric fingerprints. MaSIF \citep{gainza2020deciphering} pioneered extracting five such features from protein surfaces and used geometric CNNs for binding site and ligand pocket prediction. Follow-up work proposed continuous surface representations (SurfPro) \citep{song2024surfpro}, multi-view integration of sequence, structure, and surface properties (HoloProt) \citep{somnath2021multi}, and surface-based masked autoencoders (Surface-VQMAE) \citep{wu2024surface}.
 
Building on these insights, we propose learning a quantized surface-aware vocabulary that encodes local chemical and geometric fingerprints of surface residues—analogous to the structure-aware vocabulary in SaProt. We integrate these surface features into the GearBind framework \citep{cai2024pretrainable} and show that explicit surface context improves performance on tasks such as binding affinity prediction.

\section{Methods}
% \label{gen_inst}

\subsection{Featurization based on local context for surface residues }

Given a protein structure, our approach represents each residue based on its local neighborhood in structural space, explicitly computing its chemical and geometric properties. We created local, context-aware surface features for each residue, which are then used to encode the residue graph and train a model to learn a codebook for protein surface. Initially, we processed each protein structure using the MaSIF \citet{gainza2020deciphering}. MaSIF decomposes a surface into overlapping radial patches (which consists of three vertices) with a fixed geodesic radius, where each vertex is assigned five features: electrostatic potential (charge), hydrophobicity, hydrogen bond interaction propensity, shape index, and distance-dependent curvature. To enhance this representation, we introduced two additional geometric features: (1) the distance from the patch centroid to the residue’s $C_{\alpha}$ atom $C_\alpha->centroid $ and (2) the angle between the vectors from the patch centroid to the $C_{\alpha}$ atom and from the $C_{\alpha}$ atom to the $C_{\beta}$ atom $C_\alpha->C_\beta $. For residues lacking a side chain, we generated a virtual $C_{\beta}$ atom following the method described in FoldSeek \citet{van2022foldseek}. These two new measurements constitute the geometric features in our model. Finally, we averaged the chemical features across three vertices within a patch, resulting in a 5-dimensional chemical feature vector. Combined with the two geometric features, this yields a 7-dimensional feature vector per patch.  

We modified the MaSIF processing code to output a vertex-to-residue mapping, enabling accurate feature computation for each surface-exposed residue. For a given residue A, we first selected all patches that are within 3\r{A} from any of its atoms. As not all of the patches will be mapped to the residue (as per MaSIF calculation) we then categorized the patches into three groups: (1) \textit{Core} — patches where all three vertices map to atoms from residue A, (2) \textit{Border} — patches where at least one vertex maps to an atom from another residue, and (3) \textit{Borrowed} — patches where no vertices map to atoms from residue A. Borrowed patches are filtered out for residue A. We assigned a label to each patch: (1) \textit{High} — if its closest atoms include $\{C, O, N, S\}$ or all heavy atoms from its assigned residue, (2) \textit{Medium} — if its closest atoms include $\{C, O,N, S\}$ or all heavy atoms from neighboring residue , and (3) \textit{Low} — for patches whose closest atoms are all hydrogen atoms. We represented each residue as a graph $G = (V, E)$, where the nodes $V = \{n_i\}_{i=1:N}$ correspond to $N$ randomly sampled patches following 70\% from high, 20\% from medium and 10\% from low patches. We chose $N$ as $32$ following MaSIF \citet{gainza2020deciphering} in their tasks and chose majority of the patches that are most closely tied to the residue, followed by those that contain side-chain atoms of the neighboring residues and finally the last fraction consisting of only hydrogen atoms from its own or neighboring residue. When reconstructing the node features we weighed the nodes according to this classification. The above-mentioned 7-dimensional features are used as node features. The patch types are illustrated in Figure \ref{fig:features}.

The edges $E=\{e_{ij}|j \in \mathcal{N}_i\}_{i=1:N}$ are defined where $\mathcal{N}_i = \{j | \mathrm{dist}(n_i,n_j) < 3\textit{\r{A}}\}$ is a set of neighbors of a node $n_i$ and $\mathrm{dist}(.,.)$ is defined as the distance between the patch centroids of nodes $n_i$ and $n_j$. In addition, a virtual node has been added to the graph that is connected to every other node in the graph through its special virtual edge. This virtual node will serve a similar purpose as the [CLS] token in transformers. Since the [CLS] token has been commonly utilized to provide sentence embedding, we used the virtual node to calculate the final graph embedding and its tokenized representation. Edges are featurized according to the centroid distances: (1) \textit{Short} — where their distance is less than 1\r{A} , (2) \textit{Medium} — where their distance is between 1\r{A} and 2\r{A} , (3) \textit{Long} — where their distance is between 2\r{A} and 3\r{A} (4) \textit{Virtual} — edge between virtual node and any other node, (5) \textit{Self} — edge connecting each node to itself, and (6) \textit{No} — nodes that are not connected to each other. Therefore, there are six categories of edges. Inspired by \citep{park2022grpe}, we have used the topological relationship $\psi(i,j)$ between nodes $n_i$ and $n_j$ based on their shortest path distance with the maximum cutoff \textit{ max-hop}. Formally $\psi(i,j)$ is featurized as following: (1) \textit{Unreachable} — No connection between two nodes, (2) \textit{shortest path distance $s$} — Shortest path distance value $s \in \{0, \cdots,\textit{max-hop}\}$ between nodes $n_i$ and $n_j$, (3) \textit{Far distance} —  If the shortest path distance is greater than \textit{max-hop}, and (4) \textit{Virtual} — edge between virtual node and any other node. Therefore, there are \textit{max-hop} + 4 categories of topological relation. 

\subsection{Surface Aware Graph Encoding}

% We denote a set of nodes on the graph $\{n_i\}_{i=1:N}$ and a set of edges on the graph $\{e_{ij}|j \in \mathcal{N}_i\}_{i=1:N}$ where $N$ is the number of nodes and $\mathcal{N}_i$ is a set neighbors of a node $n_i$. Both $n_i$ and $e_{ij}$ are positive integer numbers to index the type of nodes or edges, e.g., atom numbers or bond types of a molecule. $\psi(i,j)$ denotes a function encodes topological relationship between the node $n_i$ and $n_j$. 

We developed two approaches for surface-aware graph encoding (1) \textit{naive} approach named SAGE (Surface Aware Graph Encoding) where we only reconstructed node features provided by the adjacency matrix at both encoder and decoder modules, (2) Pi-SAGE where we reconstructed both nodes and edges simultaneously. We introduced permutation-invariance property in our modeling in order to properly align reconstructed node features and adjacency matrix. Both models have similar (1) graph encoder $g_\text{enc}$, (2) finite scale quantized protein surface tokenizer ($\mathrm{FSQ}$) while their decoder $g_\mathrm{dec}$ is different. In addition, Pi-SAGE has an additional module named \textit{permuter} $p_\mathrm{perm}$ to learn the alignment between the input graph and the reconstructed one. We described each of these components in the following subsections. 

\subsubsection{Graph Encoder}

% A typical transformer architecture consists of multiple self-attention layers that uses query and a set of key pairs to first compute an attention map and then values are computed as weighted sum with the weight on the attention map to output the hidden feature for the following layer.

We used the graph transformer with learnable relative positional encoding developed by \citep{park2022grpe} that use (1) dot-product attention commonly used in transformers, (2) learnable topological relationship $\mathcal{P}_{\psi(i,j)} \in \mathbb{R}^{d_z}$ between nodes $n_i$ and $n_j$, and (3) learnable edge relationship $\mathcal{E}_{(i,j)} \in \mathbb{R}^{d_z}$ between nodes $n_i$ and $n_j$. Let us assume $x_i \in \mathbb{R}^{d_x}$ denotes the input feature of the node $n_i$ with $d_x$ as its dimension, and $z_i \in \mathbb{R}^{d_z}$ denotes the final output feature of transformer's layer with $d_z$. First, self-attention module computes query $q_i$, key $k_i$, and value $v_i$ with independent linear transformations $W^{\mathrm{query}} \in \mathbb{R}^{d_x \times d_z}$, $W^{\mathrm{key}} \in \mathbb{R}^{d_x \times d_z}$ and $W^{\mathrm{value}} \in \mathbb{R}^{d_x \times d_z}$.

\begin{equation}
   q_i=W^{\mathrm{query}}x_i , k_i=W^{\mathrm{key}}x_i \; v_i=W^{\mathrm{value}}x_i
\end{equation}

Second, topological relationship between nodes $n_i$ and $n_j$ is calculated as:

\begin{equation}
    a_{(i,j)}^{\mathrm{topology}} = q_i  \mathcal{P}_{\psi(i,j)}^{\mathrm{query}} + k_i  \mathcal{P}_{\psi(i,j)}^{\mathrm{key}}
\end{equation}

Next, edge relationship between nodes $n_i$ and $n_j$ is calculated as:

\begin{equation}
    a_{(i,j)}^{\mathrm{edge}} = q_i  \mathcal{E}_{(i,j)}^{\mathrm{query}} + k_i  \mathcal{E}_{(i,j)}^{\mathrm{key}}
\end{equation}

Finally, the overall attention map is computed summing these three terms. Attention here denotes full pairwise attention between the nodes adjusted by the graph features from the two additional matrices. 

\begin{equation}
    \begin{split}
    a_{(i,j)} &= \frac{q_i.k_j + a_{(i,j)}^{\mathrm{topology}} + a_{(i,j)}^{\mathrm{edge}}}{\sqrt{d_z}}, \\
    \hat{a}_{(i,j)} &= \frac{\mathrm{exp}(a_{(i,j)})}{\sum_{k=1}^{N}\mathrm{exp}(a_{(i,k)})}
    \end{split}
\end{equation}

The overall attention module outputs the next hidden feature by applying weighted summation on the values 
\begin{equation}
    z_i = \sum_{j=1}^{N}\hat{a}_{(i,j)}v_j
\end{equation}

% As our approach represents a residue graph by a set of patches as nodes, we adopted the GRPE paper \citep{park2022grpe} where the authors propose to use positional encoding with graphs on Transformer architecture. At a high level there are two encodings: one for graph topology (adjacency and shortest path distance) and one for edges (capturing edge types/attributes). This approach modifies the vanila self attention used in Transformer by incorporating these positional encodings directly into both attention map and value during self-attention calculation. 
The utilized learnable relative positional encoding can be seen as an alterative to linearizing graphs, thus enabling richer node-topology and node-edge interactions since it preserves structural graph information. 

\subsubsection{Surface tokenizer}
We adopted Finite Scale Quantization \citet{mentzer2023finite} to create protein surface codebook. FSQ creates a simple, fixed grid partition in a lower-dimensional space. Let us assume the FSQ's internal dimension is represented as $d_{\mathrm{FSQ}}$ and the $i^{th}$ dimension can have $\mathrm{L}_{i}$ different integers or \textit{levels}. Therefore, overall \textit{implicit} codebook size for FSQ with $\{\mathrm{L}_1,\cdots,\mathrm{L}_{d_{\mathrm{FSQ}}}\}$ can be $|\mathcal{C}| = \prod_{i=1}^{d_{\mathrm{FSQ}}} \mathrm{L}_i$. FSQ module takes in the virtual node of a residue graph from the encoder $z_{\mathrm{graph}} \in \mathbb{R}^{d_z}$, down-project the graph representation down to $d_{\mathrm{FSQ}}$ dimension through $z_{\mathrm{latent}} = \mathrm{MLP}(z_{\mathrm{graph}}) \in \mathbb{R}^{d_{\mathrm{FSQ}}}$. Then, non-differentiable online quantization step occurs for each dimension $i$ through $z_{\mathrm{FSQ},i} = \mathrm{round}(\lfloor \mathrm{L}_i/2\rfloor \mathrm{tanh}(z_{{\mathrm{latent}},i}))$. The quantization step will bound the  encoder output to  $L$ values, which is the number of dimensions of the quantizer, and then rounding to integers, leading to quantized codebook. Since $\mathrm{round(.)}$ function is a non-differential operation, straight-through estimator (STE)  \cite{bengio2013estimating} can be used to propagate gradient through $\mathrm{round\_ste}(x) = x + \mathrm{stop\_gradient}(\mathrm{round}(x) - x)$. 

\subsubsection{Permutation invariance}

The nodes in the original residue graph do not have any positional encoding and their order is arbitrary but fixed during training. Inferring this order during training allows the decoder to align the nodes which would help in the feature loss calculation. Inspired by \citet{winter2021permutation}, we added a permuter module to reconstruct the residue graph with node features and adjacency matrix in Pi-SAGE. The permuter module learns to align input and output graph through \textit{soft} alignment. Note that in Pi-SAGE, patches from residues form un-directed graphs with the adjacency matrix $\textbf{A}_\pi \in \{0,1\}^{n \times n}$ for $n \in N $ in the node order $\pi \in \Pi$ with $\Pi$ is the set of all permutations over $V$. We defined a permutation matrix $\textbf{P} \in \mathbb{R}^{N \times N}$ that reorders nodes from order $\pi$ to order $\pi'$ as $\textbf{P}_{\pi \rightarrow \pi'} = (p_{ij}) \in \{0,1\}^{n \times n}$ with $p_{ij}=1 $ if $\pi_i = \pi'_j$ and $p_{ij}=0$ otherwise. 

Input to the permuter are the node encodings $N \times \mathbb{R}^{d_z}$ obtained from the output the encoder module. We discard the virtual node at this step and do not try to reconstruct it. The permuter module has to learn how the ordering of nodes in the graph generated by the decoder model will differ from a specific node order present in the input graph. During the learning process, the decoder will learn its own canonical ordering so that, given a latent code $z_{\mathrm{latent}}$, it will always reconstruct a graph in that order. The permuter learns to transform/permute this canonical order to a given input node order. For each node $i$ of the input graph, the permuter predicts a score $s_i$ corresponding to its probability of having a low node index in the decoded graph. By sorting the input nodes indices by their assigned scores, we inferred the output node order and constructed the corresponding permutation matrix $\mathbf{P}_{\pi \rightarrow \pi'} = (p_{ij}) \in \{0,1\}^{n \times n}$ with 
 \begin{equation}
     p_{ij}= 
\begin{cases}
    1,& \text{if } j = \mathrm{argsort}(s)_i\\
    0,              & \text{else}
\end{cases}
 \end{equation}

to align input and output node order. The argsort operation being non-differentiable, the continuous relaxation of the argsort operator proposed in \citet{prillo2020softsort, grover2019stochastic} has been used as follows 
\begin{equation}
    \mathbf{P} \approx \hat{\mathbf{P}}=\mathrm{softmax}(\frac{-d(\mathrm{sort}(s)\mathbbm{1}^\top,\mathbbm{1}s^\top)}{\tau})
\end{equation}
where the softmax operator is applied row-wise, $d(x, y)$ is the $\mathrm{L}_1$-norm and  $\tau \in \mathbb{R}_+$ a temperature parameter. 

\subsubsection{Graph decoder}

Since the quantized graph encoding from the $\mathrm{FSQ}$ module is in $d_{\mathrm{FSQ}}$, we used a simple linear layer to project it back to the $d_z$ embedding that serves the purpose of node features for the decoder denoted $z_{\mathrm{dec}}$.  Inspired by \citet{winter2021permutation}, we defined sinusoidal positional embedding $\mathrm{PE} \in \mathrm{\mathbb{R}}^{N\times d_z}$ with the i-th node's embedding for k-th dimension as follows:
\begin{equation}
    \mathrm{PE}(i)_k = 
    \begin{cases}
    \mathrm{sin}(i/10000^{2k/d_z}),& \text{if } \text{k is even}\\
    \mathrm{cos}(i/10000^{2k/d_z}),              & \text{k is odd}
    \end{cases}
\end{equation}

Then we used the learned permutation matrix $\hat{\mathbf{P}}$ to reorder the positional embedding by multiplication $\mathrm{PE}_{\mathrm{update}} = \hat{\mathbf{P}} \times \mathrm{PE}$. Finally, we concatenated the node features of the decoder with the updated positional encoding and passed them to the graph decoder. The graph decoder exactly follows the graph encoder with minor differences at the final project layers:
\begin{equation}
    \begin{split}
    z_o &=  g_{\mathrm{dec}} ([z_{\mathrm{dec}} || \mathrm(PE)_{\mathrm{update}}]) \\
    \hat{m}_{\mathrm{node}} &= W_{\mathrm{node}} z_o + b_{\mathrm{node}} \\
    \hat{m}_{\mathrm{edge}} &= W_{\mathrm{edge}} z_o + b_{\mathrm{edge}} 
    \end{split}
\end{equation}

where $\hat{m}_{\mathrm{node}}$ is used to reconstruct the initial node features $m_{\mathrm{node}}$ and $\hat{m}_{\mathrm{edge}}$ is used to reconstruct the un-directed adjacency matrix $\mathbf{A}_\pi$.

\subsubsection{Losses}

Following \citep{yang2024vqgraph} we defined node and edge reconstruction as

\begin{equation}
    \begin{split}
            \mathcal{L}_{\mathrm{rec}} &= \frac{1}{N}\sum _{i=1}^{N} (1 - \frac{m_{\mathrm{node}}^T\hat{m}_{\mathrm{node}}}{{{||m_{\mathrm{node}}||.||\hat{m}_{\mathrm{node}}||}}} \\ 
            &+ ||\mathbf{A}_\pi - \sigma(\hat{m}_{\mathrm{edge}}.\hat{m}_{\mathrm{edge}}^T)||^2)
    \end{split}
\end{equation}

where $\sigma(.)$ is the sigmoid function. In
order to push the \textit{soft} permutation matrix towards a real permutation matrix (i.e. contains one 1 in every
row and column), an additional penalty term was introduced to minimize the Shannon entropy both row-wise and column-wise: 
\begin{equation}
    \mathrm{C}(\hat{\mathbf{P}}) = \sum_{i}\mathrm{H}(\bar{\mathbf{p}}_i) + \sum_{j}\mathrm{H}(\bar{\mathbf{p}}_j)
\end{equation}
with Shannon entropy $\mathrm{H}(x)= - \sum_ix_i\mathrm{log}(x_i)$ and normalized probabilities $\bar{\mathbf{p}}_i,= \frac{\hat{\mathbf{p}}_i}{\sum_j \hat{\mathbf{p}}_{i,j}}$.

The final loss would be:

\begin{equation}
    \mathcal{L} = \mathcal{L}_{\mathrm{rec}} + \lambda \mathrm{C}(\hat{\mathbf{P}})
\end{equation}

where $\lambda$ hyper-parameter would balance between main reconstruction loss and the additional penalty.

\subsection{Graph-based binding affinity prediction}

We tested the hypothesis that chemical and geometric fingerprints obtained from protein surface contain information complementary to structure to improve protein binding affinity prediction. To test it, we adopted the GearBind \citet{cai2024pretrainable} architecture based on multi-level geometric message passing network, augmented the one-hot residue features from the residue graph (obtained by pooling the atom level features of the graph after the attention step in GearBind) with the residue graph embeddings $z_{\mathrm{graph}}$ from the fine-tuned surface tokenizer and trained the augmented feature GearBind model on the SKEMPI dataset \citet{jankauskaite2019skempi}. Since the SKEMPI dataset contains both single and multiple mutations to the wild type protein complexes, we domain adapted the pre-trained surface tokenizer on these 6k proteins. This step ensured that the surface tokenizer model had seen the distribution of the mutational dataset with the protein complexes. In addition to surface tokenizer we trained 3 ESM2 \citet{verkuil2022language} models: ESM2-150M, ESM2-650M, ESM2-3B, th 3B ProstT5 model \citet{elnaggar2021prottrans}, the 3B ProstT5  \citet{heinzinger2024bilingual} model and the SaProt model \citet{su2023saprot}. 

\begin{figure*}
    \centering
    \includegraphics[width=0.5\linewidth]{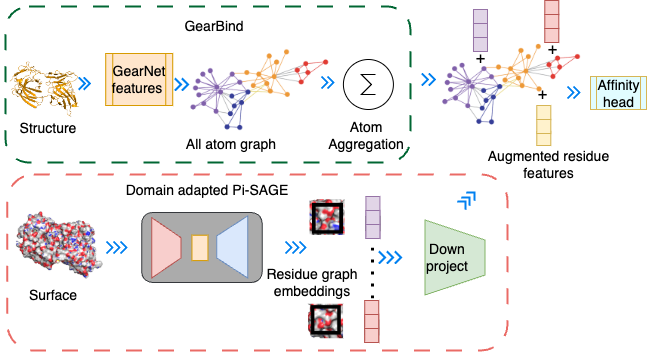}
    \caption{Training flowchart of GearBind model with domain adapted Pi-SAGE. Once the all-atom graph is constructed and reduced to the residue level graph, the one-hot residue features are augmented with surface residue embeddings to predict  $\Delta \Delta G$.}
    \label{fig:flowchart_binding}
\end{figure*}

\section{Experiments}

We performed three stage training to predict binding affinity change on SKEMPI dataset using four different sizes of surface tokenizer models with five different vocabulary sizes. In the pre-training stage, we trained on the $~200K$ experimentally validated protein structures in \citet{mentzer2023finite} RSCB dataset. We removed the SKEMPI protein complexes from the database and randomly split the proteins into $90\%$ train and $10\%$ validation splits. We used a learning rate of $2e-04$, Adam optimizer with a per GPU batch size of $32$ on a single P5 NVIDIA H200 Tensor Core GPU instances with a global batch size of $256$ for $20$ epochs. We adopted the implementation of the graph encoder module (SAGE and Pi-SAGE) from the GRPE GitHub \citet{park2022grpe}, the FSQ quantizer part from Lucidrains GitHub \citet{lucidrains} and the permutation invariant part (permuter module and the graph decoder for Pi-SAGE) from \citet{winter2021permutation}  GitHub. For SAGE the decoder module has the same architecture as the encoder module. We followed the seminal FSQ paper \citet{mentzer2023finite} to select different vocabulary sizes and hidden dimensions. For collating graphs in batches we used the DGL library \citet{wang2019deep} and followed the examples from their GitHub. We trained SAGE and Pi-SAGE separately using the same batch size, learning rate and the number of epochs.

In the second stage, we domain adapted the pre-trained models by further pre-training the tokenizers on the 6k wild type and mutated SKEMPI protein complexes from 340 unique protein complexes. Finally, we performed supervised fine tuning on the SKEMPI dataset with the GearBind model to predict binding affinity change or $\Delta \Delta G$ between wild type and mutated protein complexes. We trained GearBind using the same architecture as mentioned in the GitHub repository with $4$ geometric graph convolution layers with $128$ hidden dimension and using residual attention at the final layer. We used Rosetta \citet{rohl2004protein} to generate the mutated protein structures from the wild-type protein complexes. We obtained 3-fold cross-validation splits from the RDE paper \citet{luo2023rotamer}. The dataset is split into three folds by structure, each containing unique protein complexes that do not appear in other folds. 

We reported the average metrics across three splits. We employed five metrics to assess the accuracy of binding affinity change predictions: Pearson and Spearman correlation coefficients, root mean square error (RMSE), mean absolute error (MAE), and area Under the receiver operating characteristic curve (AUROC). For per-structure metrics, we followed the approach of \citet{luo2023rotamer} by organizing mutations according to their associated structures. Groups with fewer than ten mutation data points are excluded from this analysis. Correlation calculations are done independently for each structure, with two additional metrics: the average per-structure Pearson and Spearman correlation coefficients. Calculating AUROC involves classifying mutations according to the direction of their $\Delta \Delta $G values.  For each of the baselines (ESM, ProstT5), we augmented the node features with residue embeddings after down projection. 

\begin{table}[htbp]
\centering
\caption{Different model sizes of Pi-SAGE}
\label{tab:model_sizes}
\begin{tabular}{c|c|c|c|c}
\toprule
Model & $\#$layers & $\#$heads & hdim & $\#$params\\
\midrule
Small & 2 & 2 & 512 & 13M  \\
Medium & 4 & 4 & 768 & 44M  \\
Large & 8 & 8 & 1024 & 134M \\
XLarge & 16 & 16 & 1280 & 378M  \\
\bottomrule
\end{tabular}
\end{table}

\begin{figure*}[!htbp]
    \centering
    \includegraphics[width=0.8\linewidth]{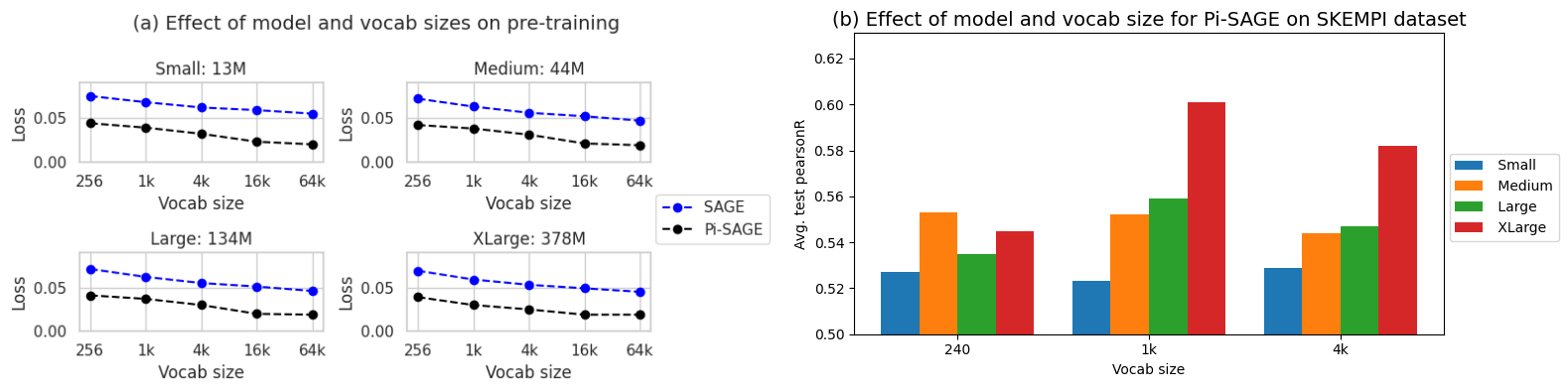}
    \caption{Effect of model and vocab sizes on pre-training and downstream task. Left: Loss curves for different model sizes and vocab sizes from pre-training the SAGE and Pi-SAGE on the RSCB database shows that with increase in model and vocab size total loss decreases. Right: Average pearson-R across three folds for different model and vocab sizes of Pi-SAGE}
    \label{fig:scaling}
\end{figure*}

\section{Results}
\subsection{Pre-training}

We pre-trained four different model sizes on five vocabulary sizes for both SAGE and Pi-SAGE (see Table \ref{tab:model_sizes}. We performed hyper-parameter optimization experiments on the CATH 4.3 dataset \citet{sillitoe2015cath} and used the learning rate, learning scheduler and weight decay from these experiments in pre-training on the RSCB database. As illustrated in Figure \ref{fig:pretrain_loss}c and \ref{fig:pretrain_loss}d, the total loss for Pi-SAGE begins higher than that of SAGE in the early stages of training but drops below SAGE’s loss as training progresses. This is because the permuter loss in training mode is quite high at the beginning and then becomes $10^{-4}$ (Figure \ref{fig:pretrain_loss}e and \ref{fig:pretrain_loss}f) leading to the lower loss for Pi-SAGE version. We hypothesize that by requiring Pi-SAGE to figure out the whole residue graph (by reconstructing both node features and the adjacency matrix), it learns to better reconstruct the node features that is demonstrated with lower feature reconstruction loss (See figure \ref{fig:pretrain_loss}a and \ref{fig:pretrain_loss}b). The feature reconstruction loss also decreases with increases in both model and vocabulary sizes (See Figure \ref{fig:scaling} left panel). This suggests that, for a fixed model size, increasing the vocabulary yields finer-grained surface representations, while scaling both the encoder and decoder enhances the model’s overall capacity.

\subsection{Binding affinity prediction}
The residue graph in GearBind is formed by pooling the atom features per-residue which contains protein structure information. We show in Table~\ref{tab:main_results} that augmenting the one-hot node features of the residue graph with residue embeddings from large sequence, structure or surface aware models improves its ability to predict change in binding affinity. This suggests that each model contains complementary information that improves the already powerful GearBind model's ability to predict $\Delta \Delta G$. For example, the large protein models like the ESM 3B model \citet{rives2019biological} ) that learn about protein structures implicitly improved the overall pearson R from $0.525$ to $0.567$ (Table \ref{tab:main_results} row 4). Structure aware models trained explicitly on protein structures like ProstT5 \citet{heinzinger2023bilingual}) improved the overall pearson R from $0.525$ to $0.545$ (Table \ref{tab:main_results} row 7). Even within the same multi-modal protein model, one modality might outperform the other. For example,  ProstT5 structure embeddings outperform ProstT5 sequence embeddings (Table \ref{tab:main_results} rows 6 and 7). Finally, as most mutations occur at the interface of two proteins for binding affinity changes, a smaller model like Pi-SAGE, trained to explicitly encode context-aware surface features of residues, consistently outperforms the larger sequence-only and sequence-structure models by improving the GearBind's prediction $0.525$ to $0.6$  on average across the test splits. We expect that Pi-SAGE performance would improve with increase in both data and model sizes.

\begin{table*}[ht]
\centering
\caption{Performance on SKEMPI dataset}
\label{tab:main_results}
\begin{small}
\begin{tabular}{l|cc|ccccc}
\toprule
\textbf{Model} & \multicolumn{2}{c|}{\textbf{Per structure}} & \multicolumn{5}{c}{\textbf{Overall}} \\
\cmidrule(r){2-3} \cmidrule(l){4-8}
 & \textbf{Pearson $\uparrow$} & \textbf{Spear. $\uparrow$} & \textbf{Pearson $\uparrow$} & \textbf{Spear. $\uparrow$} & \textbf{RMSE $\downarrow$} & \textbf{MAE $\downarrow$} & \textbf{AUROC $\uparrow$} \\
\midrule
Gearbind  & 0.365 +/- 0.082 & 0.299 +/- 0.053 & 0.525 +/- 0.106 & 0.372 +/- 0.035 & 1.921 +/- 0.277 & 1.403 +/- 0.208 & 0.650 +/- 0.006 \\
\midrule
+ ESM150M & 0.378 +/- 0.050 & 0.326 +/- 0.047 & 0.563 +/- 0.088 & 0.400 +/- 0.014 & 1.866 +/- 0.259 & 1.359 +/- 0.209 & 0.655 +/- 0.028  \\
+ ESM650M &0.381 +/- 0.063 & 0.316 +/- 0.052 & 0.539 +/- 0.096 & 0.377 +/- 0.047 & 1.852 +/- 0.226 & 1.349 +/- 0.170 & 0.652 +/- 0.032 \\
+ ESM3B & 0.418 +/- 0.088 & 0.338 +/- 0.067 & 0.567 +/- 0.057 & 0.425 +/- 0.039 & 1.834 +/- 0.144 & 1.331 +/- 0.114 & 0.671 +/- 0.026 \\
+ ProtT5 & 0.376 +/-0.112 & 0.325 +/- 0.080 & 0.551 +/- 0.088 & 0.400 +/- 0.056 & 1.873 +/- 0.179 & 1.375 +/- 0.135 & 0.665 +/- 0.019 \\
+ ProstT5 (seq) & 0.372 +/- 0.094 & 0.316 +/- 0.087 & 0.540 +/- 0.085 & 0.390 +/- 0.070 & 1.90 +/- 0.173 & 1.401 +/- 0.146 & 0.660 +/- 0.046 \\
+ ProstT5 (struct) & 0.400 +/- 0.076 & \textbf{0.347 +/- 0.049} & 0.545 +/- 0.092 & 0.408 +/- 0.032 & 1.953 +/- 0.190 & 1.436 +/- 0.137 & 0.662 +/- 0.020 \\
+ SaProt & 0.332 +/- 0.092 & 0.268 +/- 0.071 & 0.527 +/- 0.065 & 0.362 +/- 0.014 & 1.948 +/- 0.234 & 1.439 +/- 0.183 & 0.659 +/- 0.009 \\
\midrule
+ SAGE & 0.386 +/- 0.082 & 0.314 +/- 0.068 & 0.546 +/- 0.114 & 0.383 +/- 0.039 & 1.864 +/- 0.246 & 1.350 +/- 0.176 & 0.660 +/- 0.013 \\ 
+ Pi-SAGE & \textbf{0.423 +/- 0.091} & 0.345 +/- 0.077 & \textbf{0.600 +/- 0.084} & \textbf{0.428 +/- 0.038} & \textbf{1.817 +/- 0.241} & \textbf{1.306 +/- 0.200} & \textbf{0.691 +/- 0.026} \\
\bottomrule
\end{tabular}
\end{small}
\end{table*}

\section{Ablation studies}

\subsection{Effect of permutation-invariance}

The difference between SAGE and Pi-SAGE is that in the former both the encoder and decoder modules are provided the shortest path distance matrix and the edge type matrix along with the node features for the encoder. While the decoder module reconstructs the node features of the residue graph, it does not need to explicitly learn the residue graph with the node connections. On the other hand, in Pi-SAGE the encoder receives the same three matrices but the decoder needs to reconstruct both the node features in a specific order (learned by the permuter module) and the adjacency matrix containing the node connections. We hypothesize that by explicitly learning the node connections through the adjacency matrix the feature reconstruction ability of the decoder in Pi-SAGE increases. This approach allows the tokenizer to capture complex spatial relationships and biochemical properties that are crucial to understanding protein function, while ensuring that the encoded representation remains consistent regardless of arbitrary node orderings in the input graph. As we show in Table \ref{tab:main_results} both overall average metrics and per structure metrics for Pi-SAGE is better than SAGE for the same model and vocab size.

\subsection{Effect of scaling model and vocabulary sizes}
Similar to our observation in pre-training, we noticed that increasing the model size improves performance on $\Delta \Delta G$e prediction task (Figure \ref{fig:scaling}, right panel). However, we noticed that the model with 1K tokens in its vocabulary had a higher average Pearson r than both 240 and 4K tokens. We hypothesize that due to the small size of SKEMPI dataset, which has only 340 unique protein complexes (and ~6K wild type and mutated complexes), larger vocabulary might not be adding more information for the model to improve the binding affinity change (the XLarge with 1k vocab size has the higest average pearson R of 0.6 shown in  Figure \ref{fig:scaling}, right panel compared to the XLarge model with 4k vocab size with an average pearson R of 0.58). But increasing the model size might still help capture the nuances of the surface properties of the interface and improve the prediction power. 

\subsection{Effect of further pre-training on in-distribution data}
Pre-training the surface tokenizer on the RSCB database enables it to encode residue graphs of experimentally validated single chain and multi-chain proteins but it does not learn about single or multiple mutations for a protein complex. We tested whether further domain adapting the surface tokenizer on the mutated protein complexes improves the downstream task of predicting the binding affinity change. We trained GearBind with surface features from a pre-trained Pi-SAGE on RSCB dataset and showed that domain adaptation helps the model understand mutated protein complexes better than only pre-training with closed complexes (Table \ref{tab:ablation} row 1). 
\begin{table*}[ht]
\centering
\caption{Pi-SAGE ablation on SKEMPI dataset}
\label{tab:ablation}
\begin{small}
\begin{tabular}{l|cc|ccccc}
\toprule
\textbf{Pi-SAGE} & \multicolumn{2}{c|}{\textbf{Per structure}} & \multicolumn{5}{c}{\textbf{Overall}} \\
\cmidrule(r){2-3} \cmidrule(l){4-8}
 & \textbf{Pearson $\uparrow$} & \textbf{Spear. $\uparrow$} & \textbf{Pearson $\uparrow$} & \textbf{Spear. $\uparrow$} & \textbf{RMSE $\downarrow$} & \textbf{MAE $\downarrow$} & \textbf{AUCROC $\uparrow$} \\
\midrule
- Finetune  & 0.386 +/- 0.071 & 0.321 +/- 0.052 & 0.549 +/- 0.101 & 0.400 +/- 0.048 & 1.883 +/- 0.191 & 1.355 +/- 0.134 & 0.67 +/- 0.025 \\
\midrule
+ Finetune & 0.423 +/- 0.091 & 0.345 +/- 0.077 & 0.600 +/- 0.084 & 0.428 +/- 0.038 & 1.817 +/- 0.241 & 1.306 +/- 0.200 & 0.691 +/- 0.026 \\
\midrule
+ VQ  & 0.359 +/- 0.078 & 0.281 +/- 0.053 & 0.512 +/- 0.105 & 0.353 +/- 0.013 & 1.998 +/- 0.277 & 1.477 +/- 0.232 & 0.634 +/- 0.007 \\
\bottomrule
\end{tabular}
\end{small}
\end{table*}

\subsection{Effect of VQ vs. FSQ}

For a given model size, the training time for FSQ remains the same whereas for VQ it increases with increase in vocab size. For example, with VQ, time to train a 44M model for one epoch increased from 9hrs for 4k vocab to 1 day for 16k vocab on a P5 instance with 8 GPUs. Consequently, we did not train any surface tokenizer with VQ beyond a vocabulary of 4K tokens (Table \ref{tab:ablation} row 3). 

\section*{Conclusion}
Current state-of-the-art models such as GearBind is an all-atom based geometric neural network for predicting binding affinity changes between wild type and mutated protein structures. We hypothesized that explicit knowledge of surface features will improve a structure based model's ability to predict binding affinity change. We proposed Pi-SAGE, a novel approach of creating a codebook for surface exposed residues from protein structure. At its core Pi-SAGE has a graph based encoder module to encode residue graphs, a Finite Scale Quantizer to create codebook, a permuter module to learn node order of the residue graph and a decoder module to reconstruct node features and adjacency matrix. We evaluated Pi-SAGE by augmenting the residue features of GearBind to predict $\Delta \Delta G$ on SKEMPI dataset and showed that explicit knowledge of surface features improved GearBind's prediction from $0.525$ to $0.6$ on average on the test set. These results prove our hypothesis that the surface residue features from Pi-SAGE contain information above and beyond what structure can provide and boost the affinity change prediction. 

\section*{Impact Statement}

We propose Pi-SAGE, a novel surface-aware, permutation-invariant graph encoder that explicitly captures protein surface features to enhance protein binding affinity prediction. By integrating Pi-SAGE into the state-of-the-art GearBind model, we demonstrate improved accuracy in predicting $\Delta \Delta G$ between wild-type and mutant proteins. This work highlights the value of incorporating explicit surface representations in geometric deep learning models, with implications for advancing protein design, and the broader field of computational biology. We acknowledge the complexity of the method and the need for further sensitivity analysis. We plan to perform ablation studies to test the robustness of the model. 

\bibliography{example_paper}
\bibliographystyle{icml2025}

%%%%%%%%%%%%%%%%%%%%%%%%%%%%%%%%%%%%%%%%%%%%%%%%%%%%%%%%%%%%%%%%%%%%%%%%%%%%%%%
%%%%%%%%%%%%%%%%%%%%%%%%%%%%%%%%%%%%%%%%%%%%%%%%%%%%%%%%%%%%%%%%%%%%%%%%%%%%%%%
% APPENDIX
%%%%%%%%%%%%%%%%%%%%%%%%%%%%%%%%%%%%%%%%%%%%%%%%%%%%%%%%%%%%%%%%%%%%%%%%%%%%%%%
%%%%%%%%%%%%%%%%%%%%%%%%%%%%%%%%%%%%%%%%%%%%%%%%%%%%%%%%%%%%%%%%%%%%%%%%%%%%%%%
\newpage
\appendix
\onecolumn
% \section{Appendix}

\begin{figure*}
    \centering
    \includegraphics[width=0.6\linewidth]{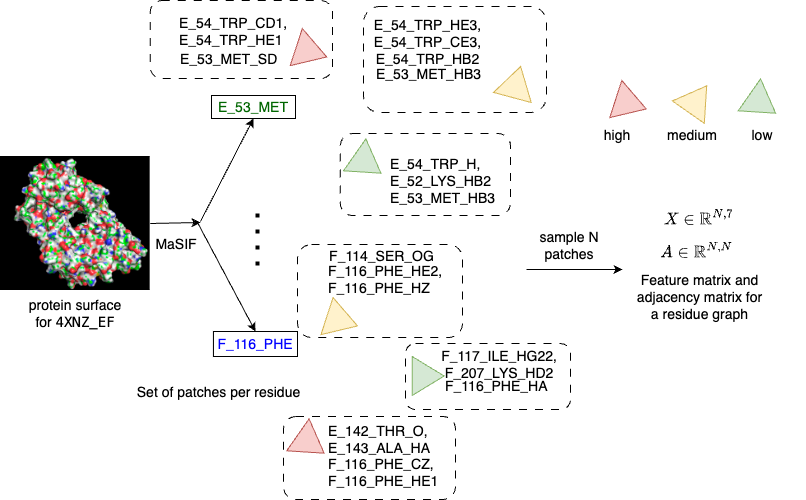}
    \caption{Creating local geometric and chemical features per residue:  Given a protein structure, we first run MaSIF and get 5 features mapped to each surface exposed residue: charge, hydrophobicity, shape index, distance dependent curvature, hydrogen bond interaction. We compute 2 more features: patch centroid to C-$\alpha$ atom of residue and angle between C-$\alpha$ to patch centroid and C-$\alpha$ to C-$\beta$. The patches are classified as high medium or low depending on the type of core or border or borrowed atoms}
    \label{fig:features}
\end{figure*}

\begin{figure*}
    \centering
    \includegraphics[width=1.0\linewidth]{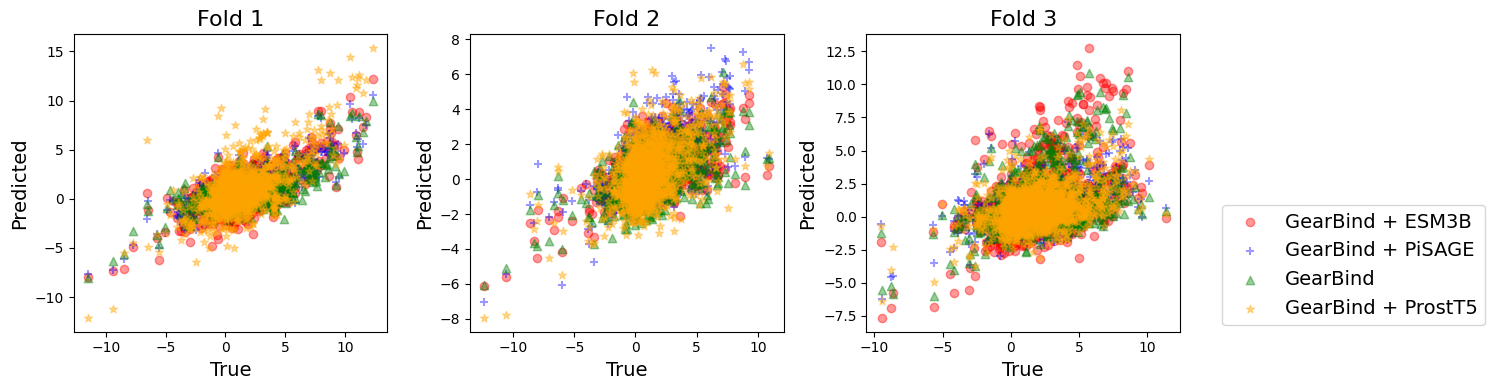}
    \caption{Pearson-R of $\Delta \Delta G$ on three folds by different methods }
    \label{fig:scatter}
\end{figure*}

\begin{figure*}
    \centering
    \includegraphics[width=0.85\linewidth]{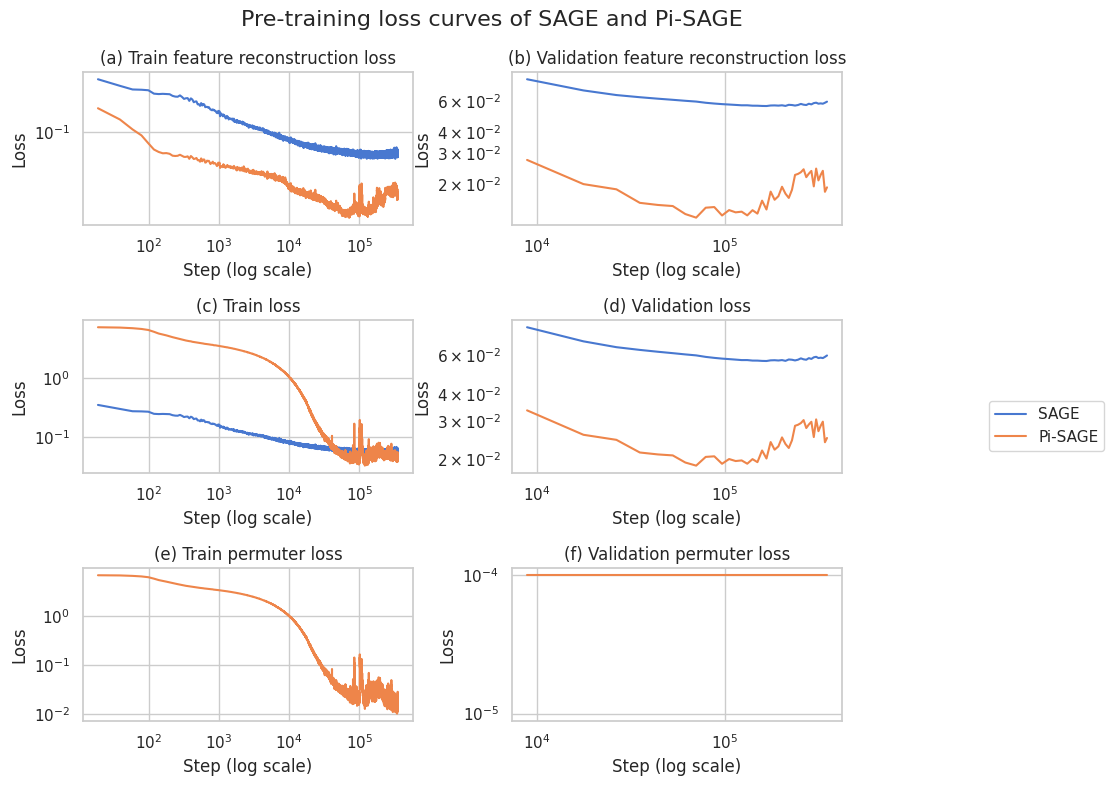}
    \caption{Pre-training loss curves for 44M 4k vocab size of SAGE and Pi-SAGE}
    \label{fig:pretrain_loss}
\end{figure*}

% The $\mathtt{\backslash onecolumn}$ command above can be kept in place if you prefer a one-column appendix, or can be removed if you prefer a two-column appendix.  Apart from this possible change, the style (font size, spacing, margins, page numbering, etc.) should be kept the same as the main body.
%%%%%%%%%%%%%%%%%%%%%%%%%%%%%%%%%%%%%%%%%%%%%%%%%%%%%%%%%%%%%%%%%%%%%%%%%%%%%%%
%%%%%%%%%%%%%%%%%%%%%%%%%%%%%%%%%%%%%%%%%%%%%%%%%%%%%%%%%%%%%%%%%%%%%%%%%%%%%%%

\end{document}